# The energy level structure of the modified Schrödinger equation can be consistent with Lamb shift

Yu-kuo Zhao[1,†], and Yu-xin Dong[2]

[1]Department of Mechanical Engineering, Shanghai Jiao Tong University, Shanghai, 200240, China

[2]School of Digital Media & Design Arts, Beijing University of Posts and Telecommunications, Beijing, 100876, China

**Abstract**

In the literature of calculating atomic and molecular structures, most Schrödinger equations are described by Coulomb potential. However, there are also a few literatures that discuss some magnetic correction methods, such as Pauli and Shortley's early work. But in fact, the calculation accuracy of these Schrödinger equations is not consistent with Lamb shift. Therefore, in the traditional ab initio calculation of quantum mechanics, it is common and necessary to use Dirac theory or quantum electrodynamics (QED) to correct the energy level of Schrödinger equation. However, the calculation of Feynman diagram is a daunting problem, including the application of self-consistent field in relativity and density functional theory. So recently, we have noticed the simplicity of the modified Newtonian mechanics, and we think that quantum mechanics will have similar properties. Here, we state this and improve the correction function in our previous action potential. In addition, through the demonstration of hydrogen-like and helium-like systems here, it can be proved that this conclusion is a potential application, that is, the energy level structure of our modified Schrödinger equation is consistent with Lamb shift.

## I. Introduction

Schrödinger equation[1-4] is fundamental to the development of modern science and has been widely used because of its maturity[5-7]. But even so, the Lamb shift[8] of hydrogen atom can't be explained by his theory—this has been known to physicists from the beginning. Therefore, a better calculation method is QED based on Dirac fine structure, and those non-relativistic approximations of relativistic quantum mechanics, please refs[9-15] to the literature. Or some kind of revision, as shown in Fig.1-3.

But in fact, in 1916, Sommerfeld was the first to explain the fine splitting of atoms as relativistic effect[16,17]. In 1921, Compton discussed the magnetism of atoms, and took the lead in proposing the quantized spin of electrons: the electron itself spinning like a tiny gyroscope[18]. However, in 1925, Uhlenbeck and Goudsmit drew physicists' attention to the discussion of Zeeman effect, Stern-Gerlach experiment and some characteristics of the old quantum theory, which can be described by the intrinsic rotating magnetic moment of electrons, thus solving some basic difficulties in the early work of Lohuizen, Sommerfeld, Landé and Pauli[19,20].

But the first relativistic wave equation is the Schrödinger-Klein-Gordon equation[21-23] proposed by Schrödinger, Klein and Gordon in some close time periods in 1926. But the more important relativistic wave equation is the Dirac equation[24,25] put forward by Dirac in 1928, and the analytical solution of the Laplace-Legendre property of single electron in Coulomb field was derived by Darwin[26] in the same year.

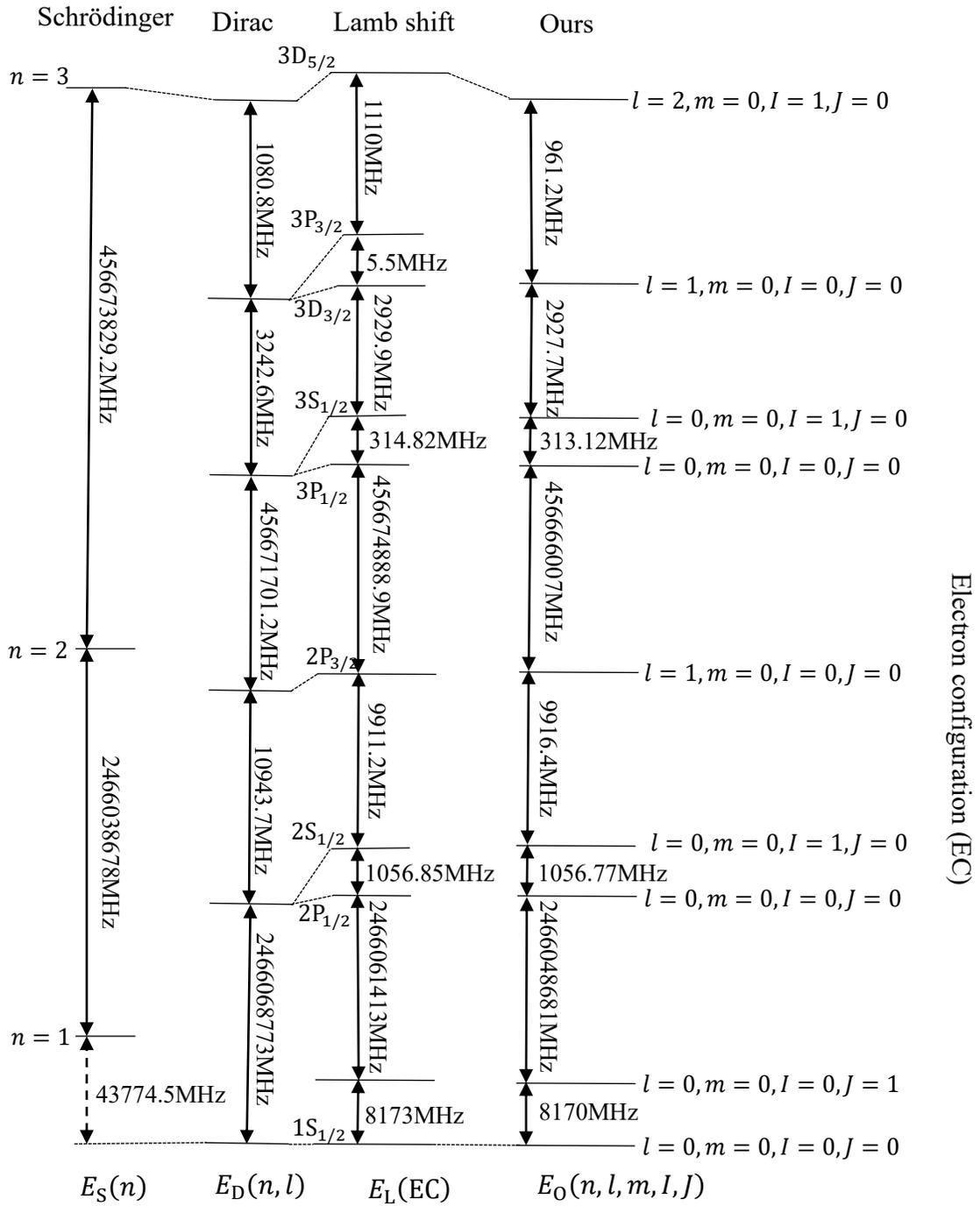

Fig.1. The energy level interval of hydrogen atom in various theoretical approximations, but excluding hyperfine structures and our redundant energy levels (Not to scale).

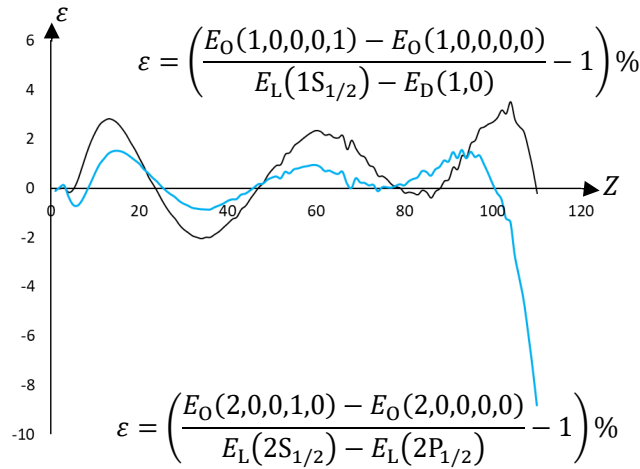

Fig.2. The error rate $\varepsilon$ of two Lamb shift in hydrogen-like system.

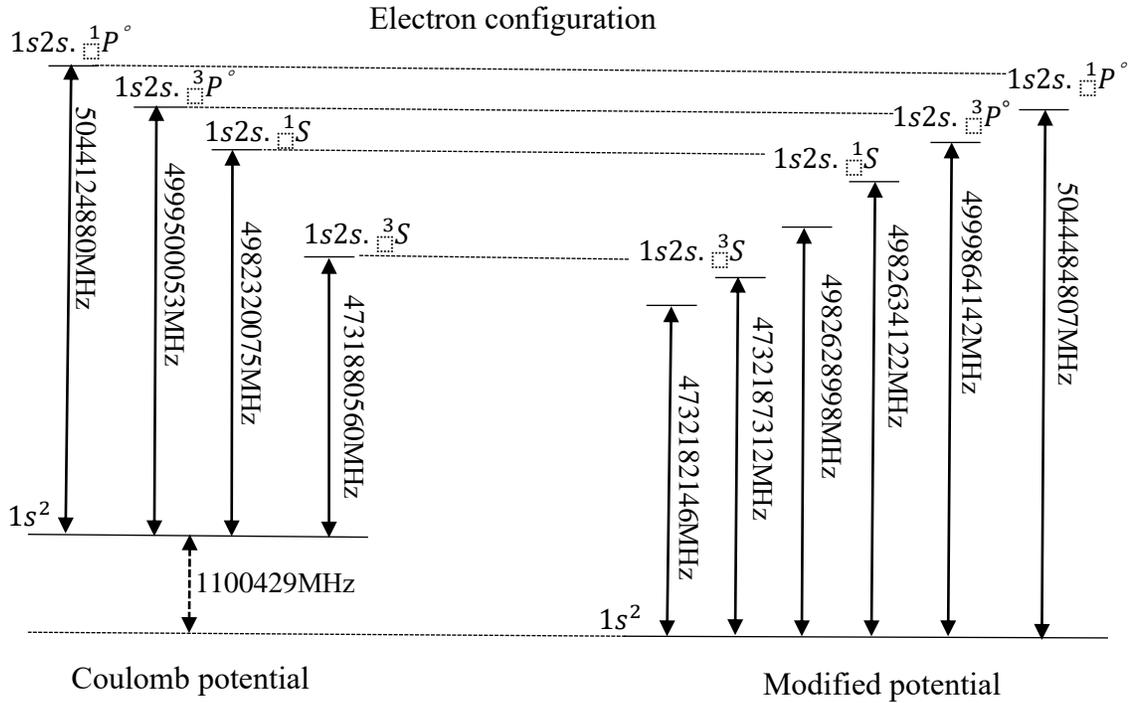

Fig.3. Lyman-$\alpha$ line of helium atom based on Coulomb potential and modified potential (Not to scale). Where, the variational method of the ground state is Hylleraas-like method, and the single excited state is Hartree-Fock method.

At the same time, a good non-relativistic approximation of Dirac equation is Pauli equation[27] proposed by Pauli in 1927, which seems to be the earliest version of the modified

Schrödinger equation. And another good discussion is the inverse-cube central force field put forward by Shortley[28] in 1931. However, the fine structure of their hydrogen atom is still not consistent with the experiment, and in 1947, Lamb and Retherford found that the $2S_{1/2}$ and $2P_{1/2}$ energy levels of hydrogen atom are not consistent with each other by means of radio frequency spectrum[8]. In other words, the Dirac equation of single electron still has some approximations. Therefore, in the same year, Bethe used the renormalization method of QED to give the calculation formula of Lamb shift under the Dirac equation of one electron[29,30], which was gradually improved by subsequent physicists, thus forming today's theoretical calculation of fine structure and ultra-fine structure with high precision, that is, Dirac-QED effect[13].

However, the complexity of Dirac equation and QED is exponential, so it will be daunting to extend them to multi-electron systems. So, in 1929, Breit constructed an approximate equation of helium-like system under non-Lorentz covariation—Breit equation[31]—he approximated the Hamiltonian of helium-like coupling under relativity as the sum of two single-electron Dirac-Hamiltonians and other operators, and interpreted these operators as the correction of magnetic interaction and delay in $v^2/c^2$ order. But in 1951, based on the Green's function of two particles, Bethe and Salpeter proposed a better substitute—Bethe-Salpeter equation[32]—and the Salpeter equation[33], Blankenbecker-Sugar-Logunov-Tavkhelidze equation[34,35], Gross equation[36] and effective Dirac equation[37] derived from these two equations became the basic approximations to deal with multi-electron bound states today, including the relativistic self-consistent field theory first proposed by Swirles[38,39]. Therefore, in traditional quantum mechanics, there is sufficient evidence to show the importance of magnetic interaction and delay effect in the fine division of hydrogen atoms. However, these evidences can't rule out Langevin's induced current hypothesis: in his view, the circular motion of electrons and nuclei will produce an induced magnetic field just like two ring circuits without resistance. As Compton said, this classical electromagnetic explanation seems to be just a matter of narrowing the error[18].

Secondly, in the experiment, our records are all probabilities in three dimensions, so there must be a fitting equation corresponding to it, such as the application of modified Newtonian mechanics in cosmology[40,41]: in some cases, the conclusion of Einstein equation can be replaced by a modified model under non-relativity. Therefore, in the one-electron Dirac equation, there must be a similar fitting force field in its inverse square law—Newton's law of universal gravitation and Coulomb's law of electrostatics have amazing mathematical similarity, so their physical mechanisms will appear in pairs. That is to say, if the spin magnetic moment and delay effect in Dirac theory are regarded as the electromagnetic action under Langevin induced current hypothesis, then the action potential of the traditional Schrödinger equation should be modified to (atomic unit):

$$V_{i,j} = \frac{\delta_0(q_i q_j)}{r_{i,j}} + \frac{\delta_1(q_i q_j)}{(x_i-x_j)^2+(y_i-y_j)^2} + \frac{\delta_2(q_i q_j)}{(z_i-z_j)^2} + \frac{\delta_3(q_i q_j)}{r_{i,j}^2} \quad (1).$$

where, the cartesian coordinates of particles is defined to be $\vec{r}_i = (x_i, y_i, z_i)$, the distance between particles is defined to be $r_{i,j} = \|\vec{r}_i - \vec{r}_j\|$, the charge of particles is defined to be $q_i$, and the modified function of action quantity is defined to be $\delta_i(q)$.

The advantage of this is that the computational complexity of fine structure and Lamb shift is obviously lower than that of Dirac and QED theory and their non-relativistic approximation, because in self-consistent field and density functional theory, we don't need to introduce Dirac operator to calculate the so-called relativistic effect, such as the demonstration of helium-like system[42,43] and the calculation accuracy shown in Fig.1-3. Meanwhile, more importantly, our correction can get a corresponding single-component wave function, which shows that there is a singularity in the density distribution of the ground-state electron cloud of hydrogen atoms—Traditional quantum mechanics holds that the ground state electron cloud of hydrogen atom is uniform on the sphere, including Dirac theory[44] (QED does not seem to have its own wave function). However, in Lamb shift's experiment, it is an indisputable fact that the energy level of ground hydrogen atom splits. Therefore, each energy level must correspond to a different electron cloud density distribution. That is to say, when the principal quantum number is 1, it will have two distribution functions, or three functions under the hyperfine structure. Therefore, our electron cloud density distribution seems to be closer to the fact, so we can draw our conclusion: through the magnetic correction of the action potential of Eq. (1), the energy level structure of the following Schrödinger equation will be consistent with Lamb shift, but lower than the hyperfine structure of QED[14], that is,

$$\mathbb{i}\frac{\partial}{\partial t}\Psi = \sum_{i\geq 0}\left(-\frac{1}{2M_i}\nabla_i^2 + \sum_{j>i}V_{i,j}\right)\Psi = E\Psi \qquad (2).$$

where, the mass of particles is defined to be $M_i$, the wave function of system is defined to be $\Psi$, the energy level of system is defined to be $E$, and the Laplacian operator is defined to be $\nabla_i^2$.

In addition, if the spin magnetic moment is the intrinsic property of electrons, the interaction between electrons and protons will not satisfy Coulomb's law at rest because there is still magnetic interaction between them. If the electron beam passes through the Stern-Gerlach experimental device, we will observe the deflection trajectory of electrons and double splitting similar to alkali metals[45], as shown in Fig.4. But in the experiment, we seem to have failed to observe these two phenomena, and the correction function of Eq. (1) is better than our previous results[46,47], as shown in the following method.

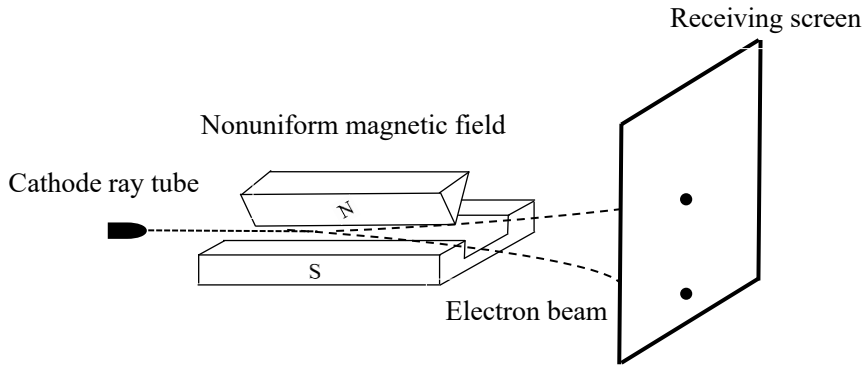

Fig. 4. Stern-Gerlach-like experiments of electron beam

## II. Methods

### A. Solution of hydrogen-like steady-state Schrödinger equation under Coulomb action potential

Under the Coulomb potential, the hydrogen-like steady-state Schrödinger equation is[1]

$$\left(-\frac{1}{2\mu}\nabla^2 - \frac{Z}{r}\right)\Psi = E\Psi \tag{3}$$

and its solution is

$$E = E_S(n) = -\frac{\mu Z^2}{2n^2} \quad \text{and} \quad \Psi = \Phi_m(\phi)\Theta_{l,m}(\theta)R_{n,l;\xi}(r) \tag{4}$$

where, the reduced mass of electrons is defined to be $\mu$, the spherical coordinate is defined to be 
$$\begin{cases} r = \sqrt{x^2 + y^2 + z^2} \\ \theta = \text{Arccos}\left(\frac{z}{r}\right) \\ \phi = \text{Arccos}\left(\frac{x}{r\sin(\theta)}\right) \end{cases}$$
, the nuclear charge number of atoms is defined to be $Z$, the principal quantum number is defined to be $n$, the angular quantum number is defined to be $l$, the magnetic quantum number is defined to be $m$, and the hydrogen-like orbital function is defined to be (non-normalized)

$$\begin{cases} \Phi_m(\phi) = \begin{cases} \cos(\phi) & \text{If}(m \geq 0) \\ \sin(\phi) & \text{Else} \end{cases} \\ \Theta_{l,m}(\theta) = \sin^{|m|}(\theta) \sum_{i=0}^{\left[\frac{l-|m|}{2}\right]} \frac{(-1)^i (2l-2i)!}{2^l i!(l-i)!(l-|m|-2i)!} \cos^{l-|m|-2i}(\theta) \quad (l \geq |m|) \\ R_{n,l;\xi}(r) = \sum_{i=0}^{n-l-1} \frac{(-1)^i}{(n-l-i-1)!(2l+i+1)!i!} (2\xi r)^{l+i} e^{-\xi r} \quad \left(n > l \text{ and } \xi = \frac{\mu Z}{n}\right) \end{cases} \tag{5}$$

### B. Darwin solution of single electron Dirac equation

Under the Coulomb potential, let the four-component wave function be denoted as $\Psi = \begin{pmatrix} \psi_1 \\ \vdots \\ \psi_4 \end{pmatrix}$, then the one-electron steady-state Dirac equation can be equivalent to[25]

$$\begin{cases} \left(E + \frac{2\mu}{\alpha^2} + \frac{\alpha Z}{r}\right)\psi_1 - \mathbb{i}\left(\frac{\partial}{\partial x} - \mathbb{i}\frac{\partial}{\partial y}\right)\psi_4 - \mathbb{i}\frac{\partial}{\partial z}\psi_3 = 0 \\ \left(E + \frac{2\mu}{\alpha^2} + \frac{\alpha Z}{r}\right)\psi_2 - \mathbb{i}\left(\frac{\partial}{\partial x} + \mathbb{i}\frac{\partial}{\partial y}\right)\psi_3 + \mathbb{i}\frac{\partial}{\partial z}\psi_4 = 0 \\ \left(E + \frac{\alpha Z}{r}\right)\psi_3 - \mathbb{i}\left(\frac{\partial}{\partial x} - \mathbb{i}\frac{\partial}{\partial y}\right)\psi_2 - \mathbb{i}\frac{\partial}{\partial z}\psi_1 = 0 \\ \left(E + \frac{\alpha Z}{r}\right)\psi_4 - \mathbb{i}\left(\frac{\partial}{\partial x} + \mathbb{i}\frac{\partial}{\partial y}\right)\psi_1 + \mathbb{i}\frac{\partial}{\partial z}\psi_2 = 0 \end{cases} \tag{6}$$

where the fine structure constant is defined to be $\alpha \approx \frac{1}{137.036}$.

Let
$$\begin{cases} \psi_1 = -\text{i} f_1(r)\Theta_{l+1,m}(\theta)e^{\text{i}m\phi} \\ \psi_2 = -\text{i} f_1(r)\Theta_{l+1,m+1}(\theta)e^{\text{i}(m+1)\phi} \\ \psi_3 = (l+m+1)f_2(r)\Theta_{l,m}(\theta)e^{\text{i}m\phi} \\ \psi_4 = (m-l)f_2(r)\Theta_{l,m+1}(\theta)e^{\text{i}(m+1)\phi} \end{cases}$$
and $\begin{cases} l \geq m \geq 0 \\ \Theta_{l,l+1}(\theta) = 0 \end{cases}$, then substitute

it into Eq.(6), and you can get:

$$\begin{cases} \left(\frac{d}{dr} - \frac{l}{r}\right)f_2(r) + \left(E + \frac{2\mu}{\alpha^2} + \frac{\alpha Z}{r}\right)f_1(r) = 0 \\ \left(\frac{d}{dr} + \frac{l+2}{r}\right)f_1(r) - \left(E + \frac{\alpha Z}{r}\right)f_2(r) = 0 \end{cases} \tag{7}$$

Let
$$\begin{cases} E = E_D(n,l) = \frac{\mu}{\alpha^2}\left(\frac{n-l-1+\sqrt{(l+1)^2-(\alpha Z)^2}}{\sqrt{(n-l-1)^2+(l+1)^2+2(n-l-1)\sqrt{(l+1)^2-(\alpha Z)^2}}} - 1\right) & (8.1) \\ \beta = \frac{n-l-1+\sqrt{(l+1)^2-(\alpha Z)^2}}{\sqrt{(n-l-1)^2+(l+1)^2+2(n-l-1)\sqrt{(l+1)^2-(\alpha Z)^2}}} & (8.2) \\ f_1(r) = \sqrt{1-\beta}\, r^{-1+\sqrt{(l+1)^2-(\alpha Z)^2}} e^{-\frac{\mu}{\alpha^2}\sqrt{1-\beta^2}\,r}(\chi_1(r) - \chi_2(r)) & (8.3) \\ f_2(r) = \sqrt{1+\beta}\, r^{-1+\sqrt{(l+1)^2-(\alpha Z)^2}} e^{-\frac{\mu}{\alpha^2}\sqrt{1-\beta^2}\,r}(\chi_1(r) + \chi_2(r)) & (8.4) \end{cases}$$

then substitute it into Eq. (7), and you can get:

$$\begin{cases} \frac{d}{dr}\chi_1(r) = \frac{(n-l-1)}{r}\chi_1(r) + \left(\frac{(l+1)}{r} + \frac{\alpha Z}{r\sqrt{1-\beta^2}}\right)\chi_2(r) \\ \frac{d}{dr}\chi_2(r) = \left(\frac{(l+1)}{r} - \frac{\alpha Z}{r\sqrt{1-\beta^2}}\right)\chi_1(r) + \left(\frac{2\mu}{\alpha^2}\sqrt{1-\beta^2} - \frac{n-l-1+2\sqrt{(l+1)^2-(\alpha Z)^2}}{r}\right)\chi_2(r) \end{cases} \tag{9}$$

Let $n = l+1 \Rightarrow \beta = \frac{1}{n}\sqrt{n^2-(\alpha Z)^2}$, then substitute it into Eq. (9), and you can get:

$$\begin{cases} \frac{d}{dr}\chi_1(r) = \frac{2n}{r}\chi_2(r) \\ \frac{d}{dr}\chi_2(r) = \left(\frac{2\mu Z}{n\alpha} - \frac{2\sqrt{n^2-(\alpha Z)^2}}{r}\right)\chi_2(r) \end{cases} \tag{10}$$

and its solution is

$$\begin{cases} \chi_1(r) = 2n\int r^{-1-2\sqrt{n^2-(\alpha Z)^2}} e^{\frac{2\mu Z}{n\alpha}r}dr \\ \chi_2(r) = r^{-2\sqrt{n^2-(\alpha Z)^2}} e^{\frac{2\mu Z}{n\alpha}r} \end{cases} \quad (Nonconvergence) \tag{11}$$

Let $n > l+1$ and $\begin{cases} \chi_1(r) = \sum_{i=0}^{n-l-1} a_i \left(\frac{2\mu}{\alpha^2}\sqrt{1-\beta^2}\,r\right)^i \\ \chi_2(r) = \sum_{i=0}^{n-l-2} b_i \left(\frac{2\mu}{\alpha^2}\sqrt{1-\beta^2}\,r\right)^i \end{cases}$, then substitute it into

Eq. (9), and you can get:

$$\begin{cases} b_i = \begin{cases} 1 & \text{If}(i=0) \\ -\dfrac{n-l-1-i}{i\left(i+2\sqrt{(l+1)^2-(\alpha Z)^2}\right)} b_{i-1} & \text{Else} \end{cases} \\ a_i = \begin{cases} \dfrac{l+1+\sqrt{(n-l-1)^2+(l+1)^2+2(n-l-1)\sqrt{(l+1)^2-(\alpha Z)^2}}}{(n-l-1)^2+2(n-l-1)\sqrt{(l+1)^2-(\alpha Z)^2}} b_{n-l-2} & \text{If}(i=n-l-1) \\ -\dfrac{l+1+\sqrt{(n-l-1)^2+(l+1)^2+2(n-l-1)\sqrt{(l+1)^2-(\alpha Z)^2}}}{n-l-1-i} b_i & \text{Else} \end{cases} \end{cases} \quad (12).$$

Therefore, the Dirac energy level of the hydrogen-like system is shown as Eq. (8.1). But its ground state wave function does not converge, as shown by Eq. (11).

### C. Energy level formula of Lamb shift (hydrogen-like)

Under the radiation correction, unit nuclear size correction, relativistic recoil correction and reduced mass correction of QED, the energy level formula of Lamb shift can be simply expressed as

$$E_L(\text{EC}) = E_D(n,l) + \frac{\alpha^3 Z^4}{n^3} \hat{F}_{n,l,j}(\alpha Z) \quad (13).$$

where the '1/2' quantum number is defined to be $j = l \pm \frac{1}{2}$, And the contribution of those corrections is represented by a dimensionless, slowly changing multivariate function $\hat{F}_{n,l,j}(\alpha Z)$—The change of the basic constant of the nucleus will affect the width of Lamb shift, such as nuclear mass, root-mean-square radius $\langle r^2 \rangle^{1/2}$ and uncertainty, and nuclear equivalent radius (homogeneously charged sphere) and uncertainty. Therefore, the explicit expression of function f is complicated, as shown in the ref[10]. So here, we fit the energy level intervals of two Lamb shifts as follows

$$\Delta_{\text{Lamb(I)}} E(Z) = E_L(1S_{1/2}) - E_D(1,0) \approx \frac{(1+0.3147(Z-1)e^{-0.776Z})}{6579683920.722}(0.000232Z^8 \\ -0.054287Z^7 + 4.954595Z^6 - 215.402047Z^5 + 5056.774393Z^4$$

$$+3330.380625Z^3) \tag{14.1}$$

$$\Delta_{\text{Lamb(II)}}E(Z) = E_L(2S_{1/2}) - E_L(2P_{1/2}) \approx \frac{(1+0.3154(Z-1)e^{-0.688Z})}{6579683920.722}(0.0000281Z^8$$

$$-0.0062514Z^7 + 0.5586651Z^6 - 24.5697836Z^5 + 639.1618597Z^4$$

$$+442.4947468Z^3) \tag{14.2}$$

### D. The solution of hydrogen-like stationary Schrödinger equation based on modified potential

Under the modification of Eq. (1), our hydrogen-like steady-state Schrödinger equation is

$$\left(-\frac{1}{2\mu}\nabla^2 + \frac{\delta_0(-Z)}{r} + \frac{\delta_1(-Z)}{x^2+y^2} + \frac{\delta_2(-Z)}{z^2} + \frac{\delta_3(-Z)}{r^2}\right)\Psi = E\Psi \tag{15}$$

So, in the spherical coordinate system, let the hydrogen-like wave function with five quantum numbers be written as

$$\Psi = \varphi_{n,l,m,I,J;\xi}(\vec{r}) = \Phi_m(\phi)\Theta_{l,m,I}(\theta)R_{n,l,m,I,J;\xi}(r) \tag{16}$$

substituting them into Eq. (15), and separating variables, we can get:

$$\begin{cases} \left(\frac{d^2}{d\phi^2} + m^2\right)\Phi_m(\phi) = 0 & (17.1) \\ \left(\frac{d^2}{d\theta^2} + \cot(\theta)\frac{d}{d\theta} - \frac{m^2+2\mu\delta_1(-Z)}{\sin^2(\theta)} - \frac{2\mu\delta_2(-Z)}{\cos^2(\theta)} + L(L+1)\right)\Theta_{l,m,I}(\theta) = 0 & (17.2) \\ \left(\frac{d^2}{dr^2} + \frac{2}{r}\frac{d}{dr} - \frac{2\mu\delta_0(-Z)}{r} - \frac{L(L+1)+2\mu\delta_3(-Z)}{r^2} + 2\mu E\right)R_{n,l,m,I,J;\xi}(r) = 0 & (17.3) \end{cases}$$

where, the angular quantum number's correction value is defined to be $L$, the newly introduced quantum number is defined to be $\begin{cases} I = \begin{cases} 0 \text{ or } 1 & \text{If}(m=0) \\ 1 & \text{Else} \end{cases} \\ J = \begin{cases} 0 \text{ or } 1 & \text{If}(l=m=I=0) \\ 0 & \text{Else} \end{cases} \end{cases}$, and the variational index is defined to be $\xi$.

So, let the non-normalized angular wave function be written as

$$\Theta_{l,m,I}(\theta) = \sin^{(-1)^{I+1}\sqrt{m^2+2\mu\delta_1(-Z)}}(\theta) \sum_{i=0}^{\left[\frac{l-|m|}{2}\right]} a_i \cos^{l-|m|+0.5-\sqrt{0.25+2\mu\delta_2(-Z)}-2i}(\theta) \quad (18).$$

and substituting them into Eq. (17.2), we can get $(a_0 = 1)$:

$$\begin{cases} L \equiv L_{l,m,I} = l - |m| + 0.5 - \sqrt{0.25 + 2\mu\delta_2(-Z)} - (-1)^I\sqrt{m^2 + 2\mu\delta_1(-Z)} \\ a_{i+1} = -\dfrac{\left(L+(-1)^I\sqrt{m^2+2\mu\delta_1(-Z)}-2i\right)\left(L+(-1)^I\sqrt{m^2+2\mu\delta_1(-Z)}-2i-1\right)-2\mu\delta_2(-Z)}{2(i+1)(2L-2i-1)} a_i \end{cases} \quad (19).$$

Let the non-normalized radial wave function be written as

$$R_{n,l,m,I,J;\xi}(r) = \sum_{i=0}^{n-l+J-1} b_i r^{i-0.5+(-1)^J\sqrt{(L+0.5)^2+2\mu\delta_3(-Z)}} e^{-\xi r} \quad (20).$$

and substituting them into Eq. (17.3), we can get:

$$\begin{cases} E = E_0(n,l,m,I,J) = -\dfrac{\mu\delta_0^2(-Z)}{2\left(n-l+J-0.5+(-1)^J\sqrt{(L+0.5)^2+2\mu\delta_3(-Z)}\right)^2} & (21.1) \\ b_{i+1} = \dfrac{\left(2i+1+(-1)^J\sqrt{(2L+1)^2+8\mu\delta_3(-Z)}\right)\xi+2\mu\delta_0(-Z)}{(i+1)\left(i+1+(-1)^J\sqrt{(2L+1)^2+8\mu\delta_3(-Z)}\right)} b_i \quad (b_0=1) & (21.2) \\ \xi = -\dfrac{\mu\delta_0(-Z)}{n-l+J-0.5+(-1)^J\sqrt{(L+0.5)^2+2\mu\delta_3(-Z)}} & (21.3) \end{cases}$$

### E. Modified function $\delta_i(q)$

Let $\begin{cases} \dfrac{\delta_0^2(q)}{2(0.5+\Lambda_1)^2} = \dfrac{1}{\alpha^2}\left(1 - \sqrt{1-(\alpha q)^2}\right) \\ \dfrac{\delta_0^2(q)}{2(1.5-\Lambda_1)^2} = \dfrac{1}{\alpha^2}\left(1 - \sqrt{1-(\alpha q)^2}\right) - \Delta_{\text{Lamb(I)}} E(|q|) \\ \dfrac{\delta_0^2(q)}{2(1.5+\Lambda_1)^2} - \dfrac{\delta_0^2(q)}{2(1.5+\Lambda_2)^2} = \Delta_{\text{Lamb(II)}} E(|q|) \\ \dfrac{\delta_0^2(q)}{2(1.5+\Lambda_1)^2} - \dfrac{\delta_0^2(q)}{2(0.5+\Lambda_3)^2} = \dfrac{1}{2\alpha^2}\left(\sqrt{4-(\alpha q)^2} - \sqrt{2+2\sqrt{1-(\alpha q)^2}}\right) \end{cases}$ , then an

effective solution of it is

$$\begin{cases} \delta_0(q) = \begin{cases} q & \text{If}(q \geq 0) \\ -\dfrac{2\sqrt{2}}{\alpha}\dfrac{\sqrt{\left(1-\sqrt{1-(\alpha q)^2}-\alpha^2\Delta_{\text{Lamb(I)}}E(|q|)\right)\left(1-\sqrt{1-(\alpha q)^2}\right)}}{\sqrt{1-\sqrt{1-(\alpha q)^2}}+\sqrt{\left(1-\sqrt{1-(\alpha q)^2}\right)-\alpha^2\Delta_{\text{Lamb(I)}}E(|q|)}} & \text{Else} \end{cases} & (22.1)\\[2em]
\Lambda_1 = \dfrac{|\delta_0(q)\alpha|}{\sqrt{2-2\sqrt{1-(\alpha q)^2}}} - 0.5 & (22.2)\\[1.5em]
\Lambda_2 = \dfrac{|\delta_0(q)(1.5+\Lambda_1)|}{\sqrt{\delta_0^2(q)-2(1.5+\Lambda_1)^2\Delta_{\text{Lamb(II)}}E(|q|)}} - 1.5 & (22.3)\\[1.5em]
\Lambda_3 = \dfrac{|\delta_0(q)(1.5+\Lambda_1)\alpha|}{\sqrt{\delta_0^2(q)\alpha^2-(1.5+\Lambda_1)^2\left(\sqrt{4-(\alpha q)^2}-\sqrt{2+2\sqrt{1-(\alpha q)^2}}\right)}} - 0.5 & (22.4)
\end{cases}$$

Let
$$\begin{cases} \left(1-\sqrt{0.25+2\delta_2(q)}-\sqrt{2\delta_1(q)}\right)^2 + 2\delta_3(q) = \Lambda_1^2 \\ \left(1-\sqrt{0.25+2\delta_2(q)}+\sqrt{2\delta_1(q)}\right)^2 + 2\delta_3(q) = \Lambda_2^2, \text{ then an effective solution} \\ \left(2-\sqrt{0.25+2\delta_2(q)}-\sqrt{2\delta_1(q)}\right)^2 + 2\delta_3(q) = \Lambda_3^2 \end{cases}$$

of it is

$$\begin{cases} \delta_1(q) = \begin{cases} 0 & \text{If}(q \geq 0) \\ \dfrac{1}{32}\left(1+\Lambda_1^2-\Lambda_3^2+\sqrt{(1+\Lambda_1^2-\Lambda_3^2)^2+4(\Lambda_2^2-\Lambda_1^2)}\right)^2 & \text{Else} \end{cases}\\[1.5em]
\delta_2(q) = \begin{cases} 0 & \text{If}(q \geq 0) \\ \dfrac{1}{64\delta_1(q)}\left(4\sqrt{2\delta_1(q)}-\Lambda_2^2+\Lambda_1^2\right)^2 - 0.125 & \text{Else} \end{cases}\\[1.5em]
\delta_3(q) = \begin{cases} 0 & \text{If}(q \geq 0) \\ 0.5\Lambda_1^2 - 0.5\left(1-\sqrt{0.25+2\delta_2(q)}-\sqrt{2\delta_1(q)}\right)^2 & \text{Else} \end{cases} \end{cases} \quad (23).$$

Therefore, the empirical formula of the electromagnetic correction function $\delta_i(q)$ in the action potential is shown in Eq. (22.1) and Eq. (23), but it does not include the nuclei outside the ref[10] (the nuclei in the non- ref[10] need the empirical data of $\Delta_{\text{Lamb(I)}}E(Z)$ and $\Delta_{\text{Lamb(II)}}E(Z)$, or are the basic constants of other nuclei, so it is not demonstrated here).

### F. Energy of helium-like ground state based on Coulomb potential (Hylleraas-like method)

Under the Coulomb potential and Born-Oppenheimer approximation, the Hamiltonian

of the helium-like steady-state Schrödinger equation is

$$\hat{H}_S = -\frac{1}{2}\nabla_1^2 - \frac{1}{2}\nabla_2^2 - \frac{Z}{r_1} - \frac{Z}{r_2} + \frac{1}{r_{1,2}} \quad (24).$$

So let the ground state variational function of Hylleraas type be[42]

$$\Psi = e^{-\xi_1 r_1 - \xi_2 r_2 + \xi_3 r_{1,2}} \quad (25).$$

and substitute it into the helium-like steady-state Schrödinger equation under Coulomb action potential, we can get:

$$\hat{H}_S|\Psi\rangle = \left(-\frac{1}{2}\xi_1^2 - \frac{1}{2}\xi_2^2 - \xi_3^2 - \frac{Z-\xi_1}{r_1} - \frac{Z-\xi_2}{r_2} - \frac{2\xi_3-1}{r_{1,2}}\right)\Psi$$

$$+ \left(\xi_1 \frac{(r_1^2 - r_2^2 + r_{1,2}^2)}{2r_1 r_{1,2}} + \xi_2 \frac{(r_2^2 - r_1^2 + r_{1,2}^2)}{2r_2 r_{1,2}}\right)\xi_3\Psi \quad (26).$$

So, its ground state energy is

$$E = E_{He-S(1S^2)} = \text{Min } E_{He-S(1S^2)}(\vec{\xi}) = -\frac{1}{2}\xi_1^2 - \frac{1}{2}\xi_2^2 - \xi_3^2$$
$$- \frac{(Z-\xi_1)F_R(\xi_1,\xi_2,\xi_3,-1,0,0) + (Z-\xi_2)F_R(\xi_1,\xi_2,\xi_3,0,-1,0)}{F_R(\xi_1,\xi_2,\xi_3,0,0,0)}$$
$$- \frac{\xi_1\xi_3 F_R(\xi_1,\xi_2,\xi_3,-1,2,-1) - \xi_1\xi_3 F_R(\xi_1,\xi_2,\xi_3,1,0,-1) - \xi_1\xi_3 F_R(\xi_1,\xi_2,\xi_3,-1,0,1)}{2F_R(\xi_1,\xi_2,\xi_3,0,0,0)}$$
$$- \frac{\xi_2\xi_3 F_R(\xi_1,\xi_2,\xi_3,2,-1,-1) - \xi_2\xi_3 F_R(\xi_1,\xi_2,\xi_3,0,1,-1) - \xi_2\xi_3 F_R(\xi_1,\xi_2,\xi_3,0,-1,1)}{2F_R(\xi_1,\xi_2,\xi_3,0,0,0)}$$
$$- \frac{(2\xi_3-1)F_R(\xi_1,\xi_2,\xi_3,0,0,-1)}{F_R(\xi_1,\xi_2,\xi_3,0,0,0)} \quad (27).$$

where, its radial coupling integral is (algorithm):

Algorithm name: $F_R(\cdot)$;

Input: $(\xi_1, \xi_2, \xi_3, k_1, k_2, k_3)$;

Output: $X = \int_0^{+\infty}\int_0^{+\infty}\int_{|r_1-r_2|}^{r_1+r_2} r_1^{k_1+1} r_2^{k_2+1} r_{1,2}^{k_3+1} e^{-2\xi_1 r_1 - 2\xi_2 r_2 + 2\xi_3 r_{1,2}} dr_1 dr_2 dr_{1,2}$;

s.t. $\begin{cases}(\xi_1, \xi_2) > 0 \text{ and } 0 < |\xi_3| < \text{Min }\{\xi_1, \xi_2\} \\ (k_1, k_2, k_3) \in \mathbb{Z} \\ k_3 + 1 \geq 0, k_2 \geq 0, \text{ and } k_1 + 2 \geq 0\end{cases}$

Algorithmic process:

$$F_R(\xi_1, \xi_2, \xi_3, k_1, k_2, k_3)\{$$

$$\text{If}(k_3 + 1 < 0)\{ \text{Return } 'Non-convergence'; \}$$

$$X \leftarrow 0; \ \text{For}(p \leftarrow 1; p < 3; p \leftarrow p + 1)\{$$

$$\text{If}(p = 2)\{v \leftarrow k_1; k_1 \leftarrow k_2; k_2 \leftarrow v; x \leftarrow \xi_1; \xi_1 \leftarrow \xi_2; \xi_2 \leftarrow x; \}$$

$$\text{For}(i_1 \leftarrow 0; i_1 \leq k_3 + 1; i_1 \leftarrow i_1 + 1)\{$$

$$\text{For}(i_2 \leftarrow 0; i_2 \leq k_3 + 1 - i_1; i_2 \leftarrow i_2 + 1)\{$$

$$\text{If}(k_2 + 1 \geq 0)\{$$

$$\text{For}(i_3 \leftarrow 0; i_3 \leq k_2 + 1 + i_2; i_3 \leftarrow i_3 + 1)\{$$

$$x \leftarrow -\frac{(-1)^{i_1+i_3}(k_3+1)!\,(k_1+k_2+k_3+4-i_1)!\,(k_2+1+i_2)!}{(k_3+1-i_1-i_2)!\,(k_2+1+i_2-i_3)!\,i_2!\,i_3!}\frac{(2\xi_1-2\xi_3)^{i_3}}{(2\xi_3)^{i_1+1}};$$

$$\text{If}(k_1 + k_3 - i_1 - i_2 + i_3 + 3 = 0)\{$$

$$X \leftarrow X - \frac{(-1)^{i_2}x}{(2\xi_2+2\xi_3)^{k_2+2+i_2}}\ln\left(\frac{\xi_1-\xi_3}{\xi_1+\xi_2}\right);$$

$$X \leftarrow X - \frac{x}{(2\xi_2-2\xi_3)^{k_2+2+i_2}}\ln\left(\frac{\xi_1+\xi_2-2\xi_3}{\xi_1-\xi_3}\right);$$

$$\} \text{ Else}\{$$

$$x \leftarrow \frac{x}{k_1+k_3-i_1-i_2+i_3+3};$$

$$X \leftarrow X + \frac{(-1)^{i_2}x}{(2\xi_2+2\xi_3)^{k_2+2+i_2}(2\xi_1-2\xi_3)^{k_1+k_3+3-i_1-i_2+i_3}};$$

$$X \leftarrow X - \frac{(-1)^{i_2}x}{(2\xi_2+2\xi_3)^{k_2+2+i_2}(2\xi_1+2\xi_2)^{k_1+k_3+3-i_1-i_2+i_3}};$$

$$X \leftarrow X + \frac{x}{(2\xi_2-2\xi_3)^{k_2+2+i_2}(2\xi_1+2\xi_2-4\xi_3)^{k_1+k_3+3-i_1-i_2+i_3}};$$

$$X \leftarrow X - \frac{x}{(2\xi_2 - 2\xi_3)^{k_2+2+i_2}(2\xi_1 - 2\xi_3)^{k_1+k_3+3-i_1-i_2+i_3}};$$

$$\}\}\} \text{ Else if}(k_2 + 2 = 0)\{$$

$$\text{For}(i_3 \leftarrow 0; i_3 \leq k_1 + k_3 + 2 - i_1 - i_2; i_3 \leftarrow i_3 + 1)\{$$

$$x \leftarrow -\frac{(-1)^{i_1}(k_3+1)!\,(k_1+k_3+2-i_1)!\,(k_1+k_3+2-i_1-i_2)!}{(k_3+1-i_1-i_2)!\,(k_1+k_3+2-i_1-i_2-i_3)!\,i_2!\,i_3!\,(2\xi_3)^{i_1+1}};$$

$$\text{If}(i_2 + i_3 = 0)\{$$

$$X \leftarrow X + \frac{x}{(2\xi_1 - 2\xi_3)^{k_1+k_3+3-i_1}} \ln\left(\frac{\xi_1 + \xi_2 - 2\xi_3}{\xi_1 + \xi_2}\right)$$

$$\} \text{ Else}\{$$

$$X \leftarrow X + \frac{(-1)^{i_2+i_3}(2\xi_2 + 2\xi_3)^{i_3}}{(2\xi_1 - 2\xi_3)^{k_1+k_3+3-i_1-i_2}(2\xi_1 + 2\xi_2)^{i_2+i_3}} \frac{x}{(i_2 + i_3)}$$

$$X \leftarrow X - \frac{(-1)^{i_3}(2\xi_2 - 2\xi_3)^{i_3}}{(2\xi_1 - 2\xi_3)^{k_1+k_3+3-i_1-i_2}(2\xi_1 + 2\xi_2 - 4\xi_3)^{i_2+i_3}} \frac{x}{i_2 + i_3}$$

$$\}\}\} \text{ Else } \{ \text{ Return } 'Non-convergence'; \}$$

$$\}\}\} \text{ Return } X; \} \qquad \text{// End.}$$

### G. Energy of helium-like ground state based on modified potential

### (Hylleraas-like method)

Under the correction of Eq. (1) and Born-Oppenheimer approximation, the Hamiltonian of our helium-like steady-state Schrödinger equation is

$$\widehat{H}_0 = \frac{1}{r_{1,2}} + \sum_{i=1}^{2}\left(-\frac{1}{2}\nabla_i^2 + \frac{\delta_0(-Z)}{r_i} + \frac{\delta_1(-Z)}{x_i^2+y_i^2} + \frac{\delta_2(-Z)}{z_i^2} + \frac{\delta_3(-Z)}{r_i^2}\right) \qquad (28).$$

So let the ground state variational function of Hylleraas type be

$$\Psi = \varphi_{1,0,0,0,0;\xi_1}(\vec{r}_1)\varphi_{1,0,0,1,0;\xi_2}(\vec{r}_2)e^{\xi_3 r_{1,2}} \qquad (29).$$

substituting it into the helium-like steady-state modified Schrödinger equation, we can get:

$$\hat{H}_0|\Psi\rangle = \left(-\frac{1}{2}\xi_1^2 - \frac{1}{2}\xi_2^2 - \xi_3^2 - \frac{\delta_0(-Z) - A_1\xi_1}{r_1} - \frac{\delta_0(-Z) - A_2\xi_2}{r_2} - \frac{2\xi_3 - 1}{r_{1,2}}\right)\Psi$$

$$+ \frac{1}{2}\left(\left(\xi_1 + \frac{1-A_1}{r_1}\right)\frac{(r_1^2 - r_2^2 + r_{1,2}^2)}{r_1 r_{1,2}} + \left(\xi_2 + \frac{1-A_2}{r_2}\right)\frac{(r_2^2 - r_1^2 + r_{1,2}^2)}{r_2 r_{1,2}}\right)\xi_3 \Psi$$

$$+ \xi_3 \sqrt{2\delta_1(-Z)}\frac{r_2}{r_1 r_{1,2}}\left(\cos(\theta_1)\cos(\theta_2) - \frac{\cos^2(\theta_1)\sin(\theta_2)\cos(\phi_1 - \phi_2)}{\sin(\theta_1)}\right)\Psi$$

$$- \xi_3 \sqrt{2\delta_1(-Z)}\frac{r_1}{r_2 r_{1,2}}\left(\cos(\theta_1)\cos(\theta_2) - \frac{\cos^2(\theta_2)\sin(\theta_1)\cos(\phi_1 - \phi_2)}{\sin(\theta_2)}\right)\Psi$$

$$+ A_3 \xi_3 \frac{r_2}{r_1 r_{1,2}}\left(\frac{\sin^2(\theta_1)\cos(\theta_2)}{\cos(\theta_1)} - \sin(\theta_1)\sin(\theta_2)\cos(\phi_1 - \phi_2)\right)\Psi$$

$$+ A_3 \xi_3 \frac{r_1}{r_2 r_{1,2}}\left(\frac{\sin^2(\theta_2)\cos(\theta_1)}{\cos(\theta_2)} - \sin(\theta_1)\sin(\theta_2)\cos(\phi_1 - \phi_2)\right)\Psi \quad (30).$$

So, our ground state energy is

$$E = E_{\text{He-O(1S}^2)} \approx \text{Min}\, E_{\text{He-O(1S}^2)}(\vec{\xi}) = -\frac{1}{2}\xi_1^2 - \frac{1}{2}\xi_2^2 - \xi_3^2$$

$$- \frac{(\delta_0(-Z) - A_1\xi_1)F_R(\xi_1, \xi_2, \xi_3, -1, 0, 0) + (\delta_0(-Z) - A_2\xi_2)F_R(\xi_1, \xi_2, \xi_3, 0, -1, 0)}{F_R(\xi_1, \xi_2, \xi_3, 0, 0, 0)}$$

$$- \frac{\xi_1\xi_3 F_R(\xi_1, \xi_2, \xi_3, -1, 2, -1) - \xi_1\xi_3 F_R(\xi_1, \xi_2, \xi_3, 1, 0, -1) - \xi_1\xi_3 F_R(\xi_1, \xi_2, \xi_3, -1, 0, 1)}{2 F_R(\xi_1, \xi_2, \xi_3, 0, 0, 0)}$$

$$- \frac{\xi_2\xi_3 F_R(\xi_1, \xi_2, \xi_3, 2, -1, -1) - \xi_2\xi_3 F_R(\xi_1, \xi_2, \xi_3, 0, 1, -1) - \xi_2\xi_3 F_R(\xi_1, \xi_2, \xi_3, 0, -1, 1)}{2 F_R(\xi_1, \xi_2, \xi_3, 0, 0, 0)}$$

$$- \frac{(1 - A_1)\big(F_R(\xi_1, \xi_2, \xi_3, -2, 2, -1) - F_R(\xi_1, \xi_2, \xi_3, 0, 0, -1) - F_R(\xi_1, \xi_2, \xi_3, -2, 0, 1)\big)\xi_3}{2 F_R(\xi_1, \xi_2, \xi_3, 0, 0, 0)}$$

$$- \frac{(1 - A_2)\big(F_R(\xi_2, \xi_1, \xi_3, -2, 2, -1) - F_R(\xi_2, \xi_1, \xi_3, 0, 0, -1) - F_R(\xi_2, \xi_1, \xi_3, -2, 0, 1)\big)\xi_3}{2 F_R(\xi_1, \xi_2, \xi_3, 0, 0, 0)}$$

$$- \frac{(2\xi_3 - 1)F_R(\xi_1, \xi_2, \xi_3, 0, 0, -1)}{F_R(\xi_1, \xi_2, \xi_3, 0, 0, 0)} \quad (31).$$

where 
$$\begin{cases} A_1 = 0.5 + \sqrt{(L_{0,0,0} + 0.5)^2 + 2\delta_3(-Z)} \\ A_2 = 0.5 + \sqrt{(L_{0,0,1} + 0.5)^2 + 2\delta_3(-Z)} \\ A_3 = 0.5 - \sqrt{0.25 + 2\delta_2(-Z)} \end{cases}.$$

**H. Energy of some helium-like singlet excited states based on Coulomb potential or**

### modified potential (Hartree-Fock method)

In the Hartree-Fock variational method of helium-like, the energies of some singlet excited states are[39,43]

$$E_{\text{He}}(\text{EC}) \approx \text{Min } E(\xi_1, \xi_2) = \begin{cases} F_0(\xi_1,\xi_2) + F_1(\xi_1,\xi_2) - F_2(\xi_1,\xi_2) & \text{If}(1s2s.\ ^3S) \\ F_0(\xi_1,\xi_2) + F_1(\xi_1,\xi_2) + F_2(\xi_1,\xi_2) & \text{Else if}(1s2s.\ ^1S) \\ F_0(\xi_1,\xi_2) + F_3(\xi_1,\xi_2) - F_4(\xi_1,\xi_2) & \text{Else if}(1s2s.\ ^3P°) \\ F_0(\xi_1,\xi_2) + F_3(\xi_1,\xi_2) + F_4(\xi_1,\xi_2) & \text{Else if}(1s2s.\ ^1P°) \end{cases} \quad (32).$$

where
$$\begin{cases} F_0(\xi_1,\xi_2) = \begin{cases} \frac{1}{2}\xi_1^2 + \frac{1}{2}\xi_2^2 - \left(\xi_1 + \frac{1}{2}\xi_2\right)Z & \text{If}(Coulomb\ potential) \\ \left(A_1 - \frac{1}{2}\right)\xi_1^2 + \frac{1}{2}(B-1)\xi_2^2 + \left(\xi_1 + \frac{1}{2}\xi_2\right)\delta_0(-Z) & \text{Else} \end{cases} \\ F_1(\xi_1,\xi_2) = \frac{1}{2}\xi_2 - \frac{(4\xi_1^3 + 3\xi_1\xi_2^2 + \xi_2^3)\xi_2^3}{2(\xi_1+\xi_2)^5} \\ F_2(\xi_1,\xi_2) = \frac{(20\xi_1^2 - 60\xi_1\xi_2 + 52\xi_2^2)\xi_1^3\xi_2^3}{3(\xi_1+\xi_2)^7} \\ F_3(\xi_1,\xi_2) = \frac{1}{2}\xi_2 - \frac{(3\xi_1+\xi_2)\xi_2^5}{2(\xi_1+\xi_2)^5} \\ F_4(\xi_1,\xi_2) = \frac{23\xi_1^3\xi_2^5}{3(\xi_1+\xi_2)^7} \\ B = 1.5 - l + \sqrt{(L_{l,0,I} + 0.5)^2 + 2\delta_3(-Z)} \end{cases} \quad (33).$$

**Funding:** No funding.
**Competing interests:** I declare have no competing interests.